\documentclass[aps,showpacs,prb,10pt,twocolumn]{revtex4-1}
\usepackage[english]{babel}
\usepackage[ansinew]{inputenc}
\usepackage{graphicx}
\usepackage{graphics}
\usepackage{amsmath}
\usepackage{amsfonts}
\usepackage{amssymb}
\usepackage{epstopdf}
\usepackage{makeidx}
\usepackage{subfigure}
\usepackage{color}
\usepackage{pgf}
\usepackage{bm}
\usepackage{tikz} 

\makeindex
\begin{document}
\title{The role of the disorder range and electronic energy in the
  graphene nanoribbons perfect transmission}
\author{Leandro R. F. Lima} \affiliation{Instituto de F\'{\i}sica,
  Universidade Federal do Rio de Janeiro, Caixa Postal 68528, Rio de
  Janeiro 21941-972, RJ, Brazil}
\author{Felipe A. Pinheiro} \affiliation{Instituto de F\'{\i}sica,
  Universidade Federal do Rio de Janeiro, Caixa Postal 68528, Rio de
  Janeiro 21941-972, RJ, Brazil}
\author{Rodrigo B. Capaz} \affiliation{Instituto de F\'{\i}sica,
  Universidade Federal do Rio de Janeiro, Caixa Postal 68528, Rio de
  Janeiro 21941-972, RJ, Brazil}
\author{Caio H. Lewenkopf} \affiliation{Instituto de F\'{i}sica,
  Universidade Federal Fluminense, 24210-346 Niter\'{o}i, Brazil}
\author{Eduardo R. Mucciolo} \affiliation{Department of Physics,
  University of Central Florida, Orlando, FL 32816-2385, USA}

\date{\today}

\begin{abstract}\noindent
 {Numerical calculations based on the recursive Green's functions
   method in the tight-binding approximation are performed to
   calculate the dimensionless conductance $g$ in disordered graphene
   nanoribbons with Gaussian scatterers. The influence of the
   transition from short- to long-ranged disorder on $g$ is studied as
   well as its effects on the formation of a perfectly conducting
   channel. We also investigate the dependence of electronic energy on
   the perfectly conducting channel. We propose and calculate a
   backscattering estimative in order to establish the connection
   between the perfectly conducting channel (with $g=1$) and the
   amount of intervalley scattering}.
\end{abstract}

\maketitle
\section{Introduction}\label{intro}

The remarkable electronic transport properties of graphene have
motivated numerous experimental and theoretical studies.
\cite{castroneto09,mucciolo10,dassarma10} Of particular interest is
the possibility of fabricating narrow width graphene samples, called
graphene nanoribbons (GNRs). By engineering the lateral confinement
one can, in principle, create an electronic energy gap leading to a
semiconductor behavior that allows for the development of novel
electronic nanodevices and applications. \cite{Geim09}

The observation of conductance quantization in GNRs turned out to be
more difficult than anticipated \cite{Lin08}. The reason is that the
vast majority of GNR samples are produced by lithographic patterning,
characterized by rough edges at the atomic scale \cite{Molitor09,
  Stampfer09, Han10, Todd09, Gallagher10}. Already at low
concentrations, such defects {can destroy} conductance quantization
\cite{Lewenkopf08}. Edge roughness may be largely suppressed in GNRs
produced by unzipping single-wall carbon nanotube. \cite{Li08, Jiao09,
  Jiao11} However, the latter is not free of bulk defects.

The way disorder affects electronic transport in GNRs strongly depends
on its spatial range. For long-ranged disorder (LRD), corresponding to
a ratio $d/a_0 \gg 1$ between the potential range $d$ and the lattice
parameter $a_0$, backscattering is suppressed and the transmission is
little affected. In contrast, short-ranged disorder (SRD) favors
scattering processes with large momentum transfer such as
backscattering. In this regime, quantum interference can cause wave
function localization. In general, these simple arguments provide a
qualitative explanation for the observed behavior of the conductance
in current experiments. Edge roughness is essentially short ranged,
while substrate impurity charges and ionically bonded adatoms are the
typical sources of long-ranged disorder. \cite{Lewenkopf08}

In view of the unavoidable disorder, the natural question that arises
is whether one can indeed observe perfect transmission or conductance
quantization in GNRs. This question has been theoretically
investigated and partially answered by Wakabayashi {\it{et al.}}
\cite{wakabayashi07,wakabayashi09}. They have found that zigzag edge
GNRs in the presence of long-range disorder exhibits a quite robust
perfectly conducting channel (PCC).

\begin{figure}[htbp]
\centering
\includegraphics[width=0.45\textwidth]{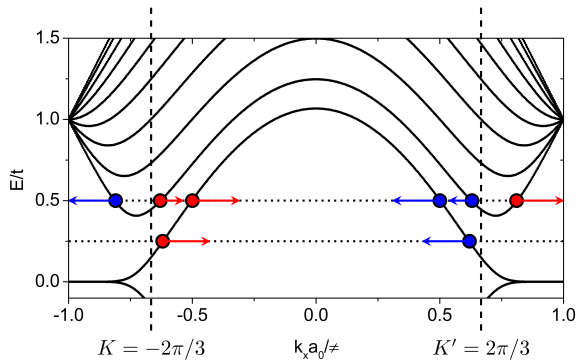}
\caption{Electronic bandstructure of a zigzag GNR with $M=10$
  horizontal chains along of width. $K$ and $K'$ are the two
  inequivalent points of the first Brillouin zone of graphene. The
  left (right) arrows indicate positive (negative) group velocity and
  forward (backwards) electronic propagation.}
\label{fig:bandstructure}
\end{figure}

The dispersion relation for zigzag GNRs, Fig.~\ref{fig:bandstructure},
helps one to understand the origin of
PCC. Fig.~\ref{fig:bandstructure} indicates two possible transversal
momentum states for electronic energies such that, at low enough
energies, only the first sub-band is allowed. The state close to the
$K$ point corresponds to a right propagating channel whereas the
other, close to $K'$, is related to a left propagating channel
\cite{wakabayashi09}. In this case, electrons propagate through the
system in a well-defined direction leading to perfect
transmission. Disorder can modify this scenario, provided it can cause
a momentum transfer $\Delta k \approx |K-K'| \approx 1/a_0$ so that it
mixes left and right propagating channels. In other words, this
backscattering process requires a momentum transfer from states at the
vicinity of the $K$-point (reminiscent of bulk graphene) to the $K'$
point, and vice-versa. The correspondence between short-ranged defects
in rough-edged GNRs and intervalley scattering has been established by
analyzing the scattering processes in both real and Fourier space
\cite{Libisch11}. For LRD such correspondence is more subtle and has
been recently addressed by analyzing the reflection probabilities of
zigzag and armchair GNRs and their symmetry properties
\cite{wurm12}. For disordered zigzag GNRs, electronic scattering
should mix valleys, leading to the suppression of transmission
depending on $d$. We address this issue by establishing a connection
between the PCC and the backscattering mechanism, analyzing both the
SRD and LRD regimes.

The purpose of this work is to determine the fundamental mechanisms
that lead to the PCC by directly analyzing the conductance dependence
on the range of disorder as the scattering potential changes from
short to long ranged. We have found that the PCC is not as robust as
previous studies pointed out. In fact, we demonstrate that the
emergence of the PCC crucially depends not only on the disorder range
but also on the electronic energy. 

This paper is organized as follows. In Sec. \ref{sec:model} we
describe the tight-binding model with on-site energy disorder, the
appearance of the PCC and the recursive Green's functions method. The
numerical results are shown in Sec.~\ref{sec:results}, where we
discuss the effect of the transition from short- to long-ranged
disorder on the conductance $g$ (and also on the PCC). We also propose
an analytical estimative for the degree of backscattering process to
elucidate the physical origin of the PCC, which is compared to the
conductance numerical results. Finally, Sec.~\ref{sec:conclusions} is
devoted to the conclusions.

\section{Model and theory}\label{sec:model}

In this Section we present the model and the theory used to address
the single-particle transport properties in GNRs. We first obtain
analytical solutions for the band structure and wave functions of the
tight-binding Hamiltonian that describes the electronic properties of
pristine zigzag GNRs. Next, we introduce the local disorder model used
to numerically study the conductance in GNRs. We close the Section
with a brief description of the numerical method employed to calculate
the transport properties, namely, the recursive Green's functions
method.

\subsection{Tight-binding model}\label{sec:tigh-binding}

\begin{figure}[htbp]
  \centering
\includegraphics[width=0.50\textwidth]{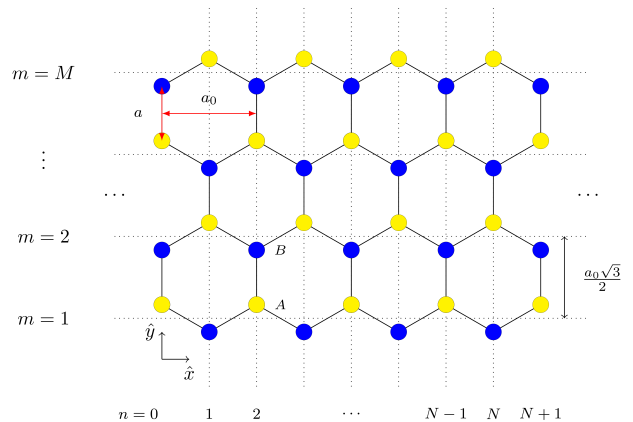}
    \caption{ (Color online) Zigzag graphene ribbon geometry. $n$ and
    $m$ label the atomic sites on the sublattices $A$ (yellow dots)
    and $B$ (blue dots). The
      atoms labeled $n=1,\cdots,N$ belong to the ribbon region and the
      atoms $n=0$ and $n=N+1$ belong to the left and right contacts,
      respectively.}
    \label{zzribbon}
\end{figure}

Close to half filling, the electrons in graphene are assumed to move
by hopping through the $p_z$ orbitals of the carbon atoms. Using the
labels introduced in Fig.~\ref{zzribbon}, the first-neighbor
tight-binding Hamiltonian for graphene reads
\begin{align}
  H = -t \sum_{n,m \in A} \bigg( 
	a^\dagger_{n,m }b^{}_{n,m+1} + 
  a^\dagger_{n ,m}b^{}_{n-1,m} + \nonumber\\ +\
	a^\dagger_{n ,m}b^{}_{n+1,m} + 
  \text{H.c.}  \bigg),
  \label{hzz1}
\end{align}
where the hopping parameter is $t=2.7\ \text{eV}$
\cite{castroneto09} and the sum is related to the sublattice $A$
sites only. The operators $a^\dagger_{n,m}\;
(b^{\dagger}_{n^\prime,m^\prime}$) create and $a^{}_{n,m}\;
(b^{}_{n^\prime,m^\prime}$) annihilate an electron at the site $(n,m)$
of the sublattice $A\, (B)$.
The integers $n=1,2$ and $m=1,\ldots,M$ label the atomic sites in the
GNR unit cell according to the notation established in
Fig.~\ref{zzribbon}. $M$ is related to the nanoribbon width by
$W=Ma_0\sqrt{3}/2$.
The lattice parameter $a_0=2.46\ \text{\r{A}}$ \cite{castroneto09}
relates to the carbon-carbon distance $a$ through $a_0=a\sqrt{3}$.

The zigzag edge boundary conditions are \cite{breyfertig06}
\begin{align}
  a^\dagger_{n,0  }|0\rangle &= 0, \label{bcam}\\
  b^\dagger_{n,M+1}|0\rangle &= 0, \label{bcbm}
\end{align}
where $|0\rangle$ is the vacuum state.
Notice that we require the states of each sublattice to vanish at
opposite edges.

The GNR eigenvalue problem reads
\begin{align}
H\left|\alpha,k_x\right> = E_\alpha(k_x) \left|\alpha,k_x\right>,
\end{align}
where $k_x$ is the longitudinal wavenumber and $\alpha$ the band (or
channel) index. Using the on-site probability amplitudes
$c_{\alpha;n,m}^{A}(k_x) = \left<0\right|a_{n,m} \left|
\alpha,k_x\right>$ and $c_{\alpha;n,m}^{B}(k_x) =
\left<0\right|b_{n,m} \left| \alpha,k_x\right>$ we write
\begin{align}
  \left| \alpha, k_x\right> = \sum_{n,m \in A}
   \left( c_{\alpha;n,m}^{A}(k_x) a^\dagger_{\alpha;n,m} +
   c_{n-1,m}^{B}(k_x) b^\dagger_{n-1,m} \right)\!\left|0\right>.
\label{psi1}
\end{align}
By inserting the eigenstates expansion (\ref{psi1}) into the
Eq.~\eqref{hzz1}, one obtains \cite{wakabayashi10}
\begin{align}
  c_{\alpha;n,m}^{A}(k_x) &= {\cal N}_\alpha(k_x) e^{ina_0k_x/2}
  \sin\!\big[ m \nu_\alpha(k_x)\big], \label{ca2}
  \\ c_{\alpha;n,m}^{B}(k_x) &= \mp {\cal N}_\alpha(k_x)
  e^{ina_0k_x/2} \sin\!\big[(m-M-1)\nu_\alpha(k_x)\big], \label{cb2}
\end{align}
where the minus and plus signs denote states with positive and
negative energies, respectively. The normalization factor
\begin{align}
  {\cal N}_\alpha(k_x) = 1/\sqrt{ N\sum_{m=1}^{M}
    \sin^2\!\big[m\nu_\alpha(k_x)\big] }
  \label{norm1}
\end{align}
is obtained by imposing $\langle \alpha, k_x|\alpha, k_x\rangle = 1$.
The momentum function $\nu_\alpha(k_x)$ is introduced to satisfy the
boundary condition (\ref{bcbm}).  $\nu_\alpha(k_x)$ is given by the
multiple solutions of the transcendental equation
\cite{wakabayashi10,breyfertig06}
\begin{align}
  2\cos(a_0k_x/2) =
  -\frac{\sin(M\nu_\alpha)}{\sin\big[(M+1)\nu_\alpha\big]}. \label{munu}
\end{align}
These analytical solutions will be used in the calculation of the
backscattering matrix elements. Finally, the zigzag GNR eigenergies
read
\begin{equation}
  E_\alpha(k_x)/t = \pm
  \left|\frac{\sin[\nu_\alpha(k_x)]}{\sin\!\big[(M+1)\nu_\alpha(k_x)\big]}\right|.
\label{enu}
\end{equation}

\subsection{Bulk disorder in GNR}\label{sec:bulk}

We calculate the conductance of disordered GNRs of length $L=Na_0/2$
(Fig.~\ref{zzribbon}). To treat disorder, with employ the Gaussian
disorder model, defined as follows. We randomly choose $N_{\rm imp}$
sites as the centers of Gaussian potentials with range $d$. $N_{\rm
  imp}$ is expressed in terms of the impurity concentration $n_{\rm
  imp}=N_{\rm imp}/N_{\rm tot}$, where the total number of atoms in
the scattering region is $N_{\rm tot}=NM$. Hence, the disorder
potential $V$ at the position ${\bf r}$ reads
\begin{align}
  V({\bf r}) = \sum_{i=1}^{N_{\rm imp}} U_i \,e^{-\left| {\bf r}-{\bf
      R}(n_i,m_i) \right|^2/d^2},
  \label{vnm}
\end{align}
where ${\bf R}(n_i,m_i)$ is the center is the $i$th Gaussian disorder
potential. The on-site lattice representation of $V$ is
\begin{align}
  V =
\sum_{n,m \in A} \left(V^{}_{n,m} a^\dagger_{n,m} a^{}_{n,m} +
V^{}_{n-1,m} b^\dagger_{n-1,m} b^{}_{n-1,m}\right),
\end{align}
where $V_{n,m}$ corresponds to $V({\bf r})$ evaluated at ${\bf
  R}(n,m)$ corresponding to the position of the lattice site $(n,m)$.

The potential amplitude $U_i$ is randomly chosen from a uniform
distribution in the interval $|U_i| \leq U_{\rm max}$, where
\begin{align}
  U_{\rm max} = \frac{2}{\sqrt{3}}U_0/\left(\sum^{\text{full
      space}}_{\bf R} e^{-{\bf R}^2/d^2}\right) .
  \label{um}
\end{align}
The dimensionless parameter $U_0/t$ defines the maximum disorder
potential energy at each impurity site.

\subsection{Recursive Green's function technique}\label{sec:RGF}

The conductance is obtained by using the recursive Green's functions
method \cite{Fisher81,Kinnon85}. This method provides a
computationally efficient way to calculate the total Green's function
of a GNR connected to pristine semi-infinite graphene leads at both
ends. Using a decimation method we compute the surface Green's
functions and the decay width functions of the left $\Gamma_0$ and
right $\Gamma_{N+1}$ leads. Next, we split the GNR ``central" region
(of width $W$ and length $L$) in $N$ slices containing $M$ transversal
sites [see Fig.\ref{zzribbon}] and iteratively calculate the total
retarded Green's function $G^r_{1,N}$ that contains {information about
  electron propagation} from slice $1$ to $N$. Finally, the
dimensionless conductance $g$ is obtained from the Caroli formula
\cite{caroli71}, namely, $g = \text{Tr} \left[ \Gamma_0 G^r_{1,N}
  \Gamma_{N+1} G^a_{1,N}\right]$. The linear electronic conductance is
$G=(2e^2/h)g$, where the factor 2 is due to the spin degeneracy.

\section{Results}
\label{sec:results}

In this Section we study the robustness of the PCC in disordered
zigzag GNRs. This is done by numerically computing the dimensionless
conductance $g$ and interpreting the results in terms of the
analytical tools presented in the previous Section.

We compute the dimensionless conductance averaged over a large number
of disorder realizations $\langle g \rangle$ (typically $10^3$) by
means of the recursive Green's function method, The impurity potential
strength is $U_0/t=1$. As in Ref.~\onlinecite{wakabayashi07}, we consider
GNRs with $M=10$. For a zigzag GNR of this width, there is a single
propagating channel for $E<E_{2} \equiv 0.406t$, where we denote the
threshold energy to open the $\alpha$th channel by $E_\alpha$.

\begin{figure}[htb]
  \centering
  \includegraphics[width=0.5\textwidth]{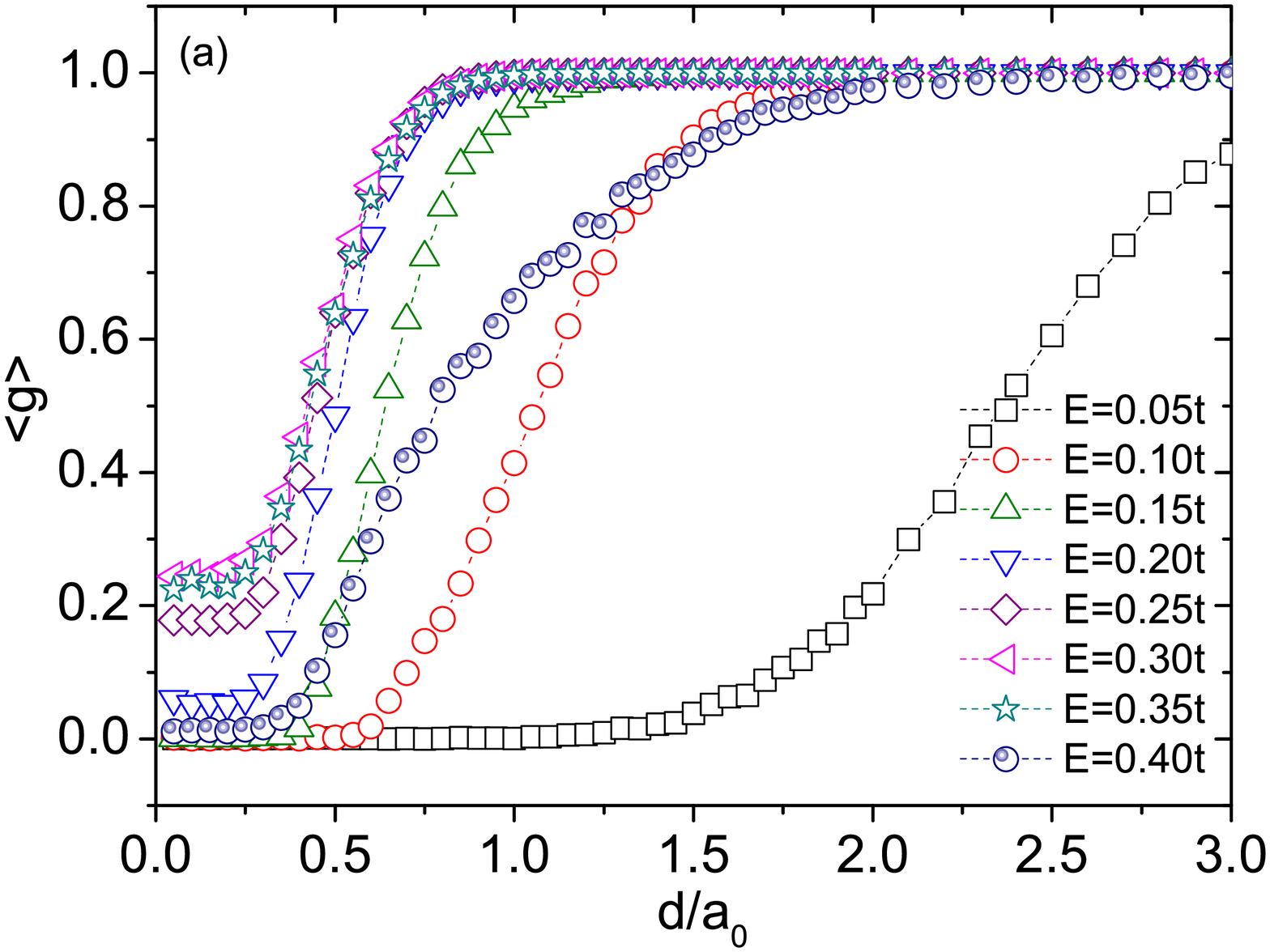}
  \includegraphics[width=0.5\textwidth]{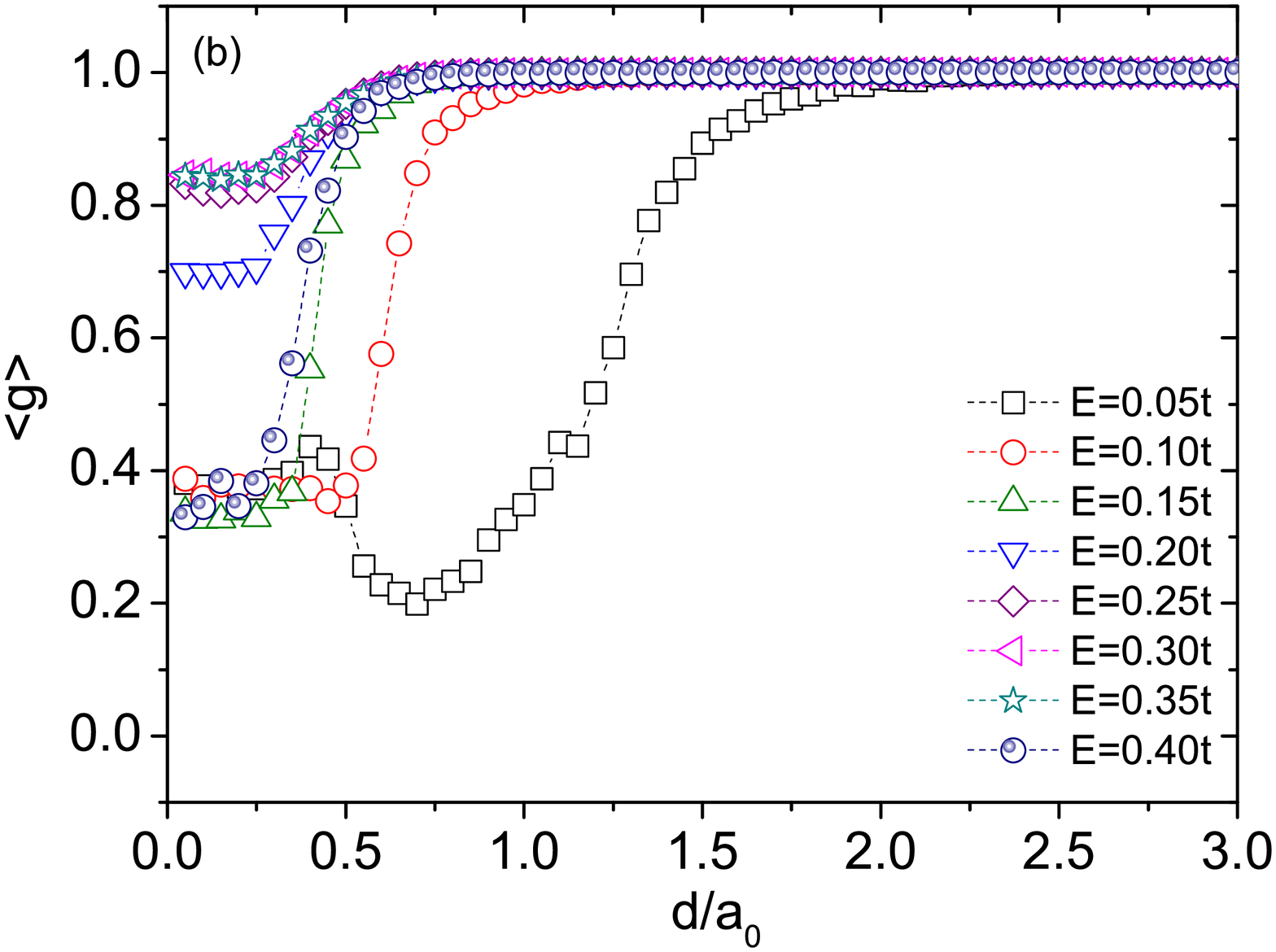}
  \caption{Average dimensionless conductance $\langle g \rangle$ as a
    function of the potential range $d$ for a ribbon of length $L =
    500a_0$ and width $W=5\sqrt{3}a_0$ for the impurity concentrations
    (a) $n_{\rm imp}=0.10$ and (b) $n_{\rm imp}=0.01$.}
  \label{gxd_frac_imp}
\end{figure}

We start by numerically investigating the behavior of the PCC in the
crossover from SRD to LRD regimes as a function of the energy $E$.
Figure~\ref{gxd_frac_imp} shows our results for $\langle g \rangle$ as
a function of the potential range $d/a_0$ for two values of impurity
concentration, namely, $n_{\rm imp}=0.10$ and $n_{\rm imp}=0.01$. The
simulations show that for sufficiently large values of $d/a_0$,
irrespective of $E$ and $n_{\rm imp}$, the PCC always occurs, as
$\langle g \rangle = 1$ within the numerical precision. Moreover, we
find that the potential range $d_{c}$, defined as the potential range
above which the PCC appears, depends strongly and non-monotonically on
energy $E$: (i) $d_c$ decreases for increasing energies $E$ starting
from the charge neutrality point, and (ii) $d_c$ increases with
increasing energy as $E$ approaches $E_{2}$. This is in contrast to
Ref.~\onlinecite{wakabayashi09}, which suggests the emergence of the PCC for
all energies $E<E_2$ at $d/a_0=1.5$ and $n_{\rm imp}=0.10$. The same
qualitative trend is found for both low ($n_{\rm imp}=0.01$) and high
($n_{\rm imp}=0.10$) impurity concentrations we analyze, as shown in
Figs.~\ref{gxd_frac_imp}(b) and \ref{gxd_frac_imp}(a),
respectively.

The overall values of the average conductance $\langle g \rangle$ are
larger in the case of more diluted impurities, as expected. For both
impurity concentrations, Fig.~\ref{gxd_frac_imp} shows conductance
plateaus for short scattering potential ranges, typically $d/a_0 \alt
0.3$. These plateaus have a simple interpretation. Let us consider a
single Gaussian disorder scattering center, placed at a site $i$. For
$d$ smaller than roughly half the inter-atomic distance $a/2 =
a_0/(2\sqrt{3}) = 0.29a_0$, the neighboring sites of $i$ are hardly
affected by the scattering center placed at $i$. Further reduction in
the potential range does not change the system Hamiltonian.  

\begin{figure}[htbp]
	\centering
		\includegraphics[width=0.50\textwidth]{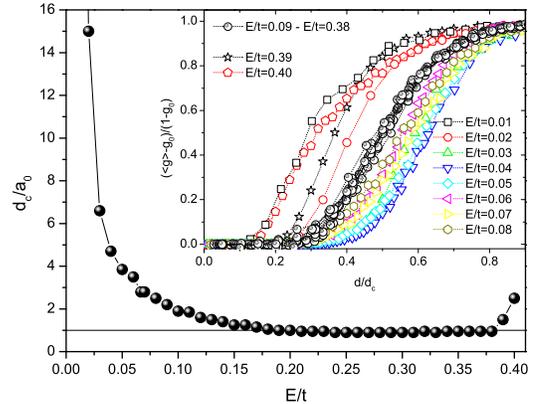}
	\caption{Disorder range above which the PCC appears, $d_c/a_0$, as a function of the electron energy $E$. Inset: Scaled average conductance $(\langle g \rangle - g_0)/(1 - g_0)$ as a function of $d/d_c$ for different energies $E/t$. }
	\label{fig:dc_manual_inset}
\end{figure}

To investigate the role of the disorder potential range on the emergence of the PCC, in Fig.~\ref{fig:dc_manual_inset}  we show $d_c/a_0$, defined above, as a function of energy $E/t$ for $n_{\rm imp}=0.10$.  Figure \ref{fig:dc_manual_inset}  shows that $d_c/a_0 \approx 1$, except for low energies and energies at the threshold of the $n=2$ channel opening. 
The insert of Fig.~\ref{fig:dc_manual_inset} indicates that the behavior of the function $\langle g  (d/a_0)\rangle$ does not show, in general, a simple single-parameter scaling behavior. The numerical results suggests that a single-parameter scaling for  $(\langle g \rangle - g_0)/(1 - g_0)$ as a function of $d/d_c$ holds approximately true only for $0.15 \alt E/t \alt 0.37$, where $g_0$ is the average conductance minimum for a fixed energy $E/t$.

To understand the numerical results presented above we investigate the
backscattering mechanisms induced by the disorder potential. These
mechanisms can be quantified by studying the backscattering matrix
elements connecting forward- to backward-moving states with
$\alpha=1$, namely,
\begin{widetext}
\begin{align}
  \left\langle {1,-k_x}\right|V\left| {1, +k_x}\right\rangle = \sum_{n,m \in A}
  \bigg\{
  \left[c_{1;n  ,m}^A(-k_x)\right]^* c_{1;n  ,m}^A(k_x) V_{n  ,m}  + 
  \left[c_{1;n-1,m}^B(-k_x)\right]^* c_{1;n-1,m}^B(k_x) V_{n-1,m}
  \bigg\},
  \label{back-scattering}
\end{align}
\end{widetext}
where the symbols are defined in Sec.~\ref{sec:model}. The expression
for the backscattering matrix element (\ref{back-scattering}) becomes
very simple in the SRD regime: For a case of a single impurity
{placed} at ${\bf R}_{n_0,m_0}$, it reads
\begin{align}
  \left\langle 1,-k_x(E)\right|V\left|1,+k_x(E)\right\rangle = \hspace{3.0cm} \nonumber\\
   =\left[ c_{n_0,m_0}^\Lambda(-k_x) \right]^* c_{n_0,m_0}^\Lambda(k_x) V_{n_0,m_0}, 
  \label{BS1}
\end{align}
where $\Lambda = A$ or $B$.

\begin{figure}[htb]
  \centering
  \includegraphics[width=0.5\textwidth]{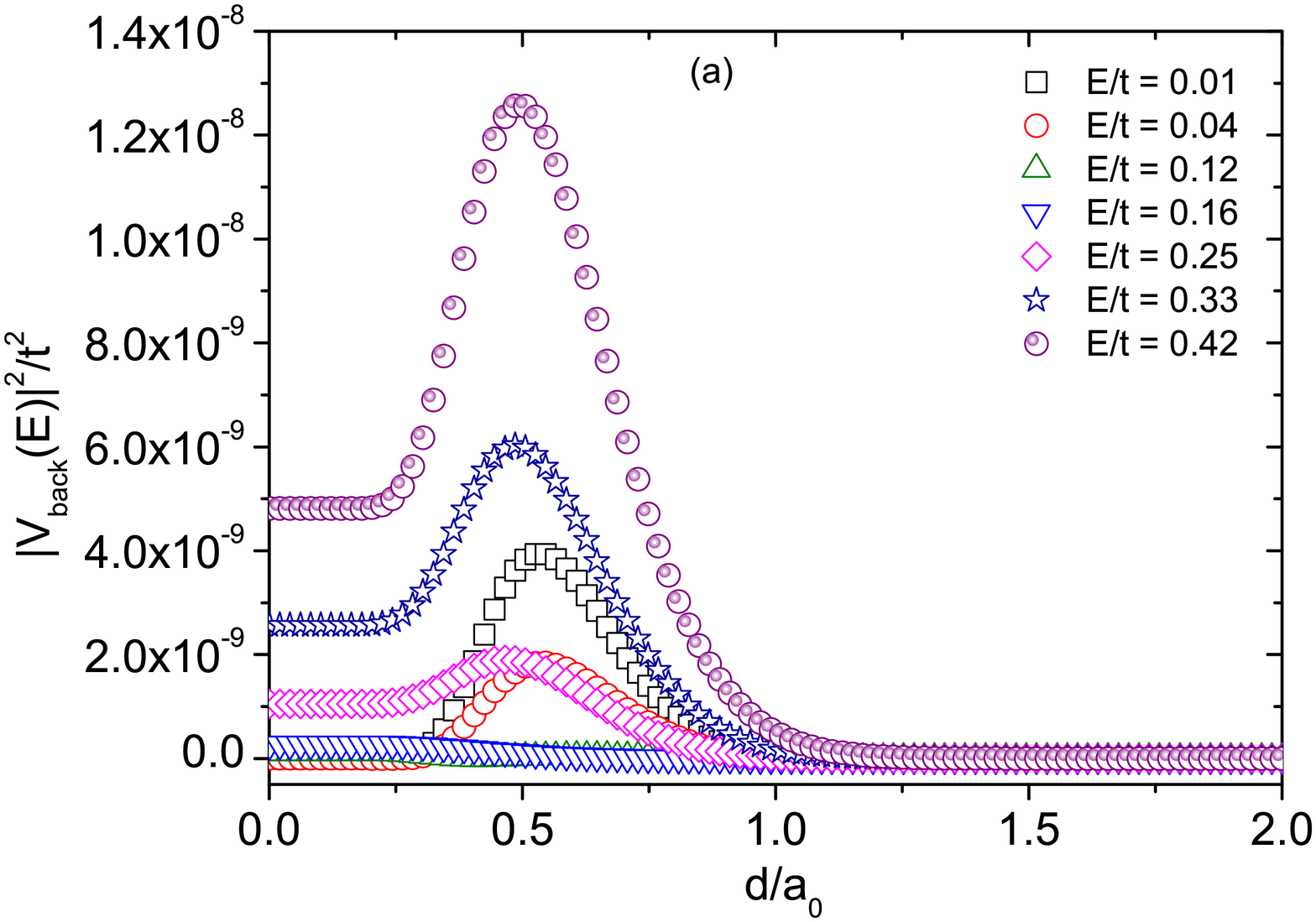}
  \includegraphics[width=0.5\textwidth]{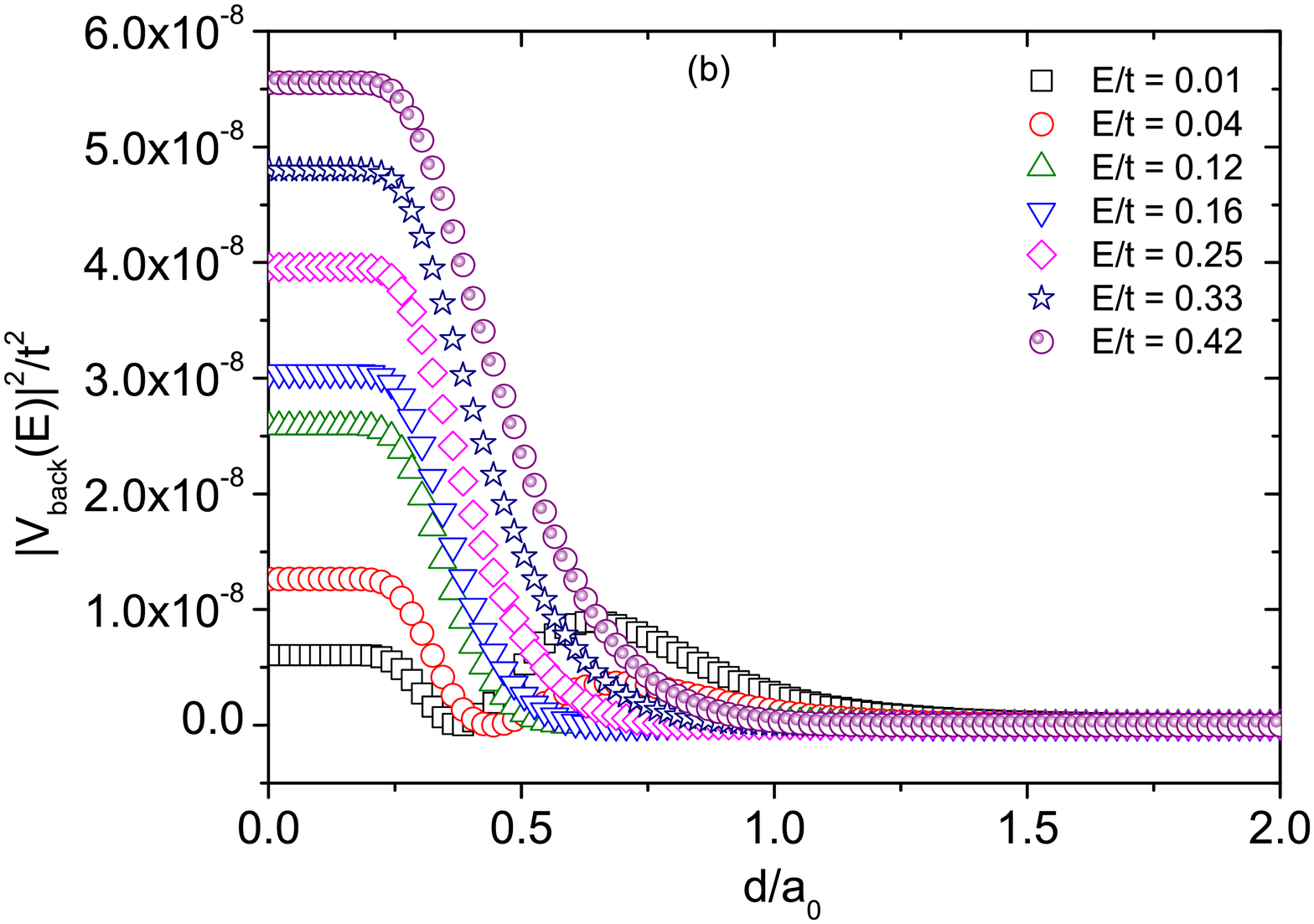}
  \caption{
    Backscattering matrix element $V_{\rm back}$ calculated as
    a function of the scattering range $d/a_0$ for different values of
    the energy $E$ for a single Gaussian impurity center placed (a) at
    the center of the GNR ($n_0=100,m_0=5$) and (b) at its edge
    vicinity ($n_0=100,m_0=9$). The GNR has the dimensions $N=200$ and
    $M=10$. For this value of GNR width, the edge state appears for
    $E<0.1t$. \cite{wakabayashi09,breyfertig06}}
  \label{fig:scatt12_imp100}
\end{figure}

Let us examine the backscattering matrix elements in a number of
representative situations. We first consider the single-impurity
scattering case. Figure \ref{fig:scatt12_imp100} shows the
backscattering matrix, Eq.~\eqref{BS1}, as a function of the energy
for two different configurations: The Gaussian scattering potential is
placed at the center or at the edge of the GNR. As for {narrow zigzag GNRs and} small values of
$E$ the electronic states are typically concentrated at the edges of
the ribbon, we expect a distinct behavior in these two limiting
cases. 

Figure \ref{fig:scatt12_imp100}(a) shows $V_{\rm back}(E) \equiv
\left\langle 1,-k_x(E)\right|V\left|1,+k_x(E)\right\rangle $ as a
function of $d/a_0$ for a single Gaussian potential placed at the
center of the GNR. Due the SRD to LRD crossover, the backscattering
matrix element decreases very fast as $d/a_0 \agt 1$. As expected from
Eq.~\eqref{BS1} and the previous discussion, $V_{\rm back}(E)$ hardly
changes for $d/a_0 \alt 0.25$. For $M=10$, the states $|1,\pm
k_x\rangle$ become localized at the GNR edges when $E/t\alt 0.1$
\cite{breyfertig06}. In this situation, impurities located at the GNR
center result in weak backscattering so that the plateaus quickly drop
to $V_{\rm back}(E)\approx 0$. Figure \ref{fig:scatt12_imp100}(b) shows
$|V_{\rm back}(E)|$ for the case of edge impurities. For $E/t >0.1$,
corresponding to ordinary states, the behavior is similar to
Fig.~\ref{fig:scatt12_imp100}(a). However, for edge states the
situation changes dramatically: Here $V_{\rm back}(E)$ decreases
surprisingly slowly with increasing $d/a_0 \agt 1$. This indicates
that, in the presence of edge states, intervalley scattering is highly
sensitive to the impurity position, even in the LRD case.

\begin{figure}[htb]
  \centering
  \includegraphics[width=0.5\textwidth]{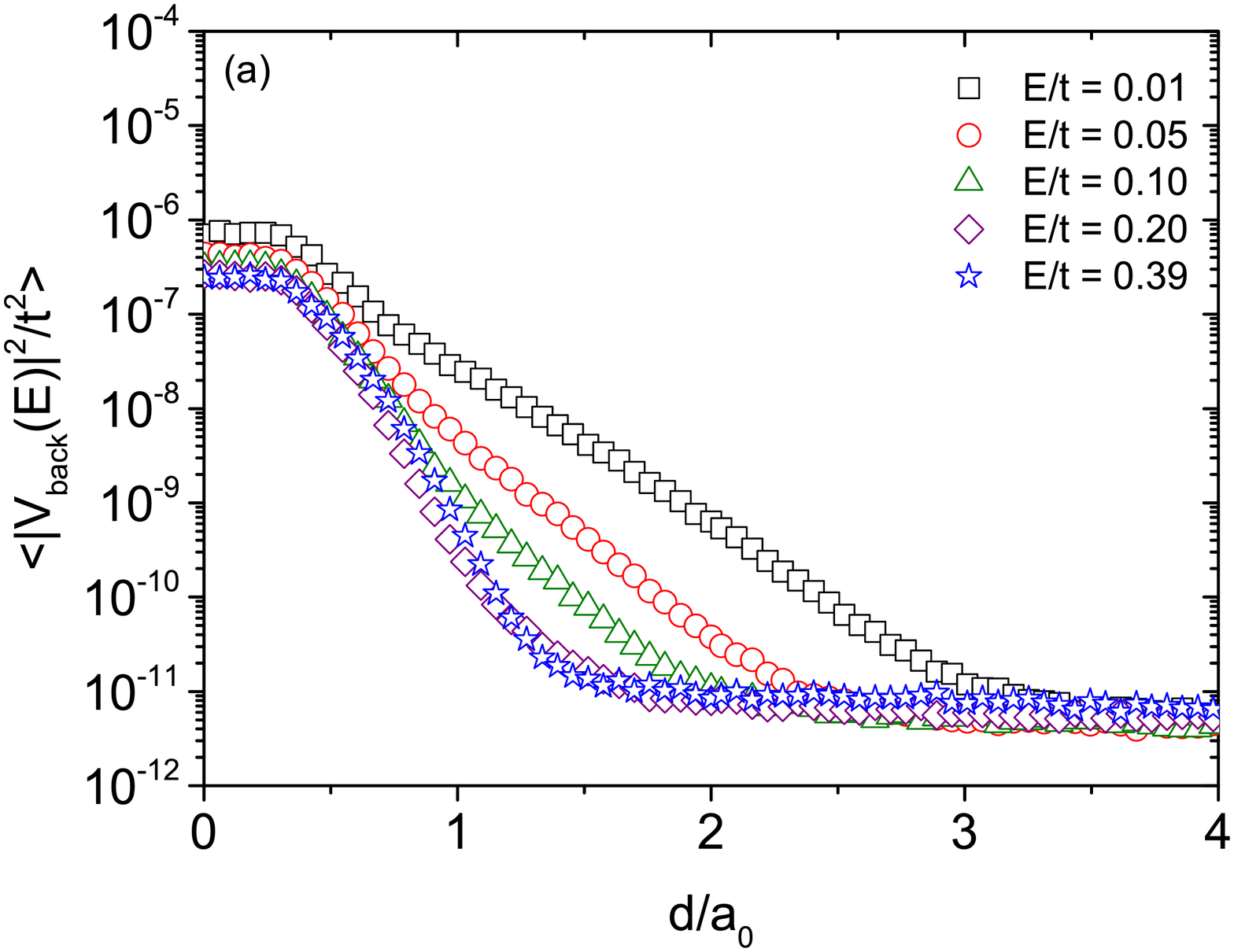}
  \includegraphics[width=0.5\textwidth]{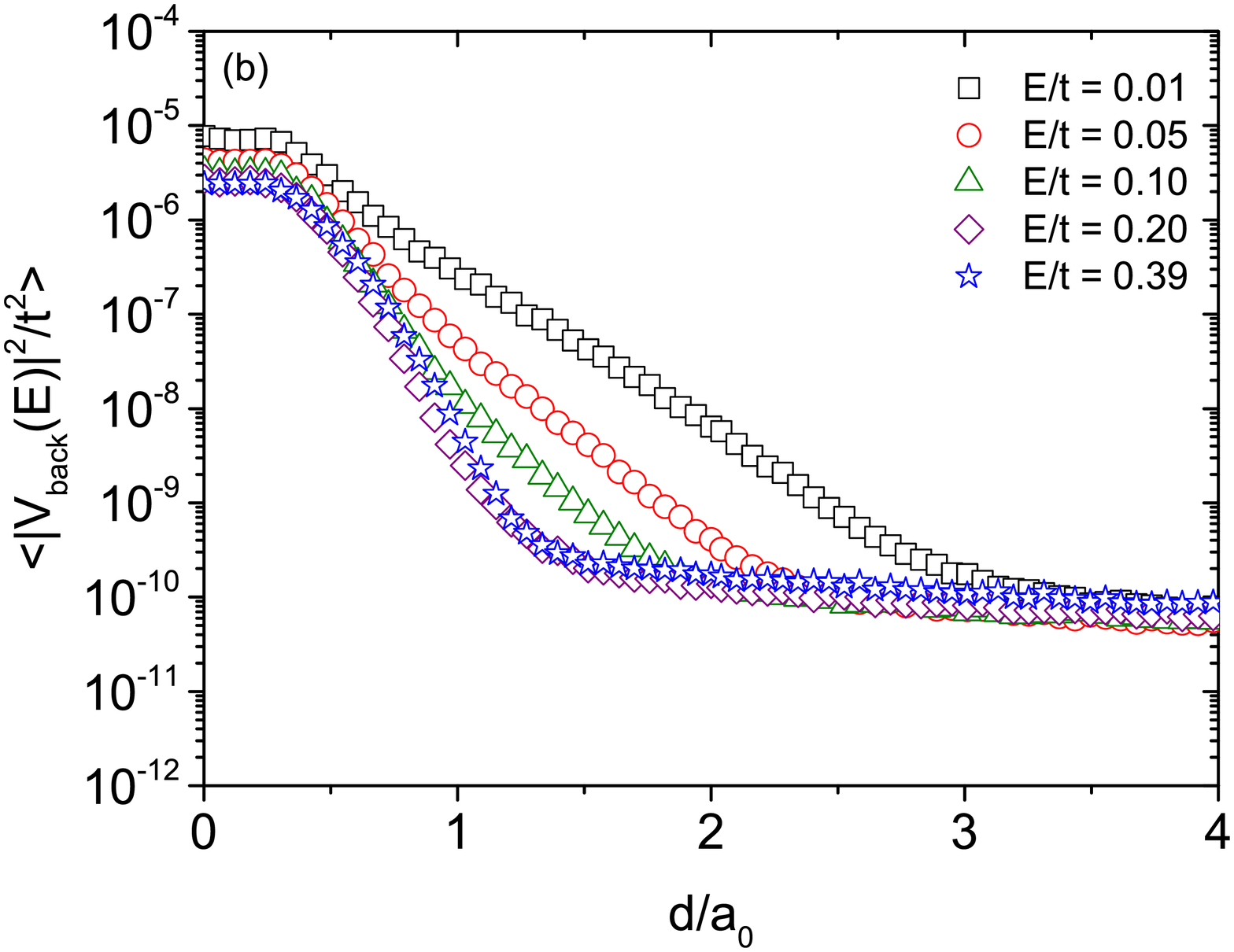}
  \caption{ 
Average backscattering matrix element $\langle |V_{\rm back}|^2
\rangle$ as a function of the disorder range $d/a_0$ for a number of
energies $E$. The average is taken over $10^3$ disorder configurations
for an impurity concentration of (a) $n_{imp}=0.01$ and (b)
$n_{imp}=0.10$. The GNR has the dimensions $L=500a_0$ and
$W=10(\sqrt{3}/2)a_0$. This ribbon presents edge state for $E<0.1t$.}
  \label{fig:-kVk_x_d_frac_imp}
\end{figure}

We now consider the situations of low ($n_{imp}=0.01$) and large ($n_{imp}=0.10$) impurity concentrations, the same values used in the conductance numerical calculations (Fig.~\ref{gxd_frac_imp}). We compute the backscattering matrix element using Eq.~(\ref{back-scattering}) and their average over different configurations of disorder. The results are shown in Fig.~\ref{fig:-kVk_x_d_frac_imp} and are contrasted with the average conductance $\langle g \rangle$. Like $\langle g \rangle$, the average backscattering matrix elements, $\langle | V_{\rm back}(E)|^2\rangle$, also display a plateau-like behavior for $d/a_0<0.3$. For $|E|>0.15t$, $\langle | V_{\rm back}(E)|^2\rangle$ exhibits a fast decay with increasing $d/a_0$ independent of the energy $E$. On the other hand, for $|E|<0.15t$ the backscattering matrix elements decay nearly exponentially for $1.0\alt d/a_0\alt 2.0$, reaching minimum values depending on $n_{\rm imp}$. For energies near the Dirac point ($E=0$), backscattering is maximal, persisting even in the LRD regime. In summary, in the SRD regime the averaged backscattering reaches its maximum value causing a conductance suppression. In distinction, in general $\langle | V_{\rm  back}(E)|^2\rangle$ is suppressed in the LRD favoring the {appearance of the} PCC provided $d/a_0\agt 1$. 

The small energy regime is an exception to this picture: 
The enhanced backscattering near the charge neutrality point can be understood in simple terms. Qualitatively, one expects that a scattering potential with a characteristic length scale $d$ can only effectively backscatter electron states with initial momentum $k$ to final momentum $-k$  provided $1/d \approx |\!-2k+G|$, where $G$ is a reciprocal lattice vector. Nonzero $G$ vectors describe Umklapp scattering processes. As one approaches the charge neutrality point, scattering between $k\approx\pi/a$ to $-k\approx-\pi/a$ mediated by a $G=2\pi/a$ Umklapp becomes dominant, so that $|\!-2k+G|$ approaches zero and even a very smooth potential (large $d$) is able to scatter the electron states and destroys PCC. That is the reason for the sharp increase of $d_c$ near $E=0$, shown in Fig. \ref{fig:dc_manual_inset}.

We discuss now the robustness of the PCC for energies $E$ approaching
{$E_2$}. As discussed, for such energies $V_{\rm back} \approx 0$ as
long as $d/a_0 \agt 1$. To explain deviations from the PCC,
first-order perturbation theory is not sufficient since LRD cannot
account for the large momentum transfer necessary for backscattering,
as already discussed. LRD can only account for backscattering
processes in the vicinities of the energy {$E_2$}, where left and
right propagating modes {(Fig.~\ref{fig:bandstructure})} are not far away in
momentum space. For a sufficiently smooth long-range disorder, one can
account for disorder effects by introducing a local chemical potential
as $\mu[V({\bf r})]$. In this scenario, LRD can suppress the PCC at
energies {$E_2 - \delta E \alt E \alt E_2+ \delta E$}, where $\delta E$
is the typical disorder potential fluctuation.

We have also carried out numerical simulations for higher energies (up to $E/t=0.9$), corresponding to the multi-channel case. However, in this case we have not found any qualitative difference between our results and the ones reported in Ref.~\onlinecite{wakabayashi09}. More precisely, we have found deviations from the PCC for incident energies at the vicinity of energies corresponding to the crossover between $ g  = 2n-1$ and $2n +1$ for clean ribbons ($n$ being the number of channels). This behavior corroborates the results of Ref.~\onlinecite{wakabayashi09}. 

\section{Conclusions}\label{sec:conclusions}

In summary, the electronic conductance of disordered GNRs in general,
and the emergence of the PCC in particular, present a richer behavior
than expected from previous studies. Specifically, the critical
impurity potential $d_c$ for the PCC emergence shows a strong and
clear dependence on the electronic energy. This result, obtained
numerically from recursive Green's functions calculations, is
explained in simple physical grounds by calculations of energy and
potential range dependences of electron intervalley scattering
probabilities, which follows the opposite qualitative trends of the
conductance. This occurs for sufficiently low energies such that the
Fermi level crosses just one band (single channel conductance). This
behavior confirms and justifies the simple picture of an electron
undergoing intervalley scattering only if the impurity potential has
substantial Fourier components at large momenta, thus being able to
provide a momentum transfer $\Delta k \approx 1/a_0$ to the
electron. In other words, for a single channel, backscattering occurs
only if intervalley scattering does. This picture not only explains
the conductance behavior for low energies, but it also provides us
with guidelines to explain the more complicated cases in which the
Fermi level crosses many channels. In that case, as shown in Fig. 1,
backscattering (i.e., group velocity reversal) can occur even for
small momentum transfers (intravalley scattering), therefore
explaining the disappearance of the PCC as the Fermi level approaches
the threshold for opening the second transmission channel.

\begin{acknowledgments}
	This work is supported by Brazilian funding agencies CAPES, CNPq, FAPERJ and INCT - Nanomateriais de Carbono.
\end{acknowledgments}

\bibliography{PCC-1025}

\end{document}